\documentclass[showpacs,superscriptaddress,prl,twocolumn]{revtex4}
\usepackage{amsmath,amssymb,amsfonts,amsthm}

\newtheorem{theorem}{Theorem}

\newcommand{\mf}{\mathcal{M}}

\newcommand{\dv}{dS_0}

\newcommand{\dvs}{dS_{\Sigma}}

\newcommand{\Lsig}{\Delta_{\Sigma}}

\newcommand{\Rs}{R_{\Sigma}}

\begin{document}
\title{Area -- Angular momentum inequality for axisymmetric black
holes}

\author{Sergio Dain}
\email[E-mail: ]{dain@famaf.unc.edu.ar}
\affiliation{Facultad de Matem\'atica, Astronom\'{\i}a y
  F\'{i}sica, FaMAF, Universidad
  Nacional de C\'ordoba,\\
  Instituto de F\'{\i}sica Enrique Gaviola, IFEG, CONICET,\\
  Ciudad Universitaria (5000) C\'ordoba, Argentina} 
\affiliation{Max Planck
  Institute for Gravitational Physics (Albert Einstein Institute) Am
  M\"uhlenberg 1 D-14476 Potsdam Germany.}

\author{Martin Reiris}
\email[Email: ]{martin@aei.mpg.de}
\affiliation{Max Planck
  Institute for Gravitational Physics (Albert Einstein Institute) Am
  M\"uhlenberg 1 D-14476 Potsdam Germany.}

\date{\today}

\begin{abstract}  
 We prove the local inequality $A \geq 8\pi|J|$, where $A$ and $J$ are the area
 and angular momentum of any axially symmetric closed stable minimal surface in
 an axially symmetric maximal initial data. From this theorem it is proved that
 the inequality is satisfied for any surface on  complete asymptotically flat
 maximal axisymmetric data. In particular it holds for marginal or event
 horizons of black holes. Hence, we prove the validity of this inequality for
 all dynamical (not necessarily near equilibrium) axially symmetric black
 holes. 
\end{abstract}


\maketitle

\emph{Introduction.}  --- Black holes in
equilibrium are characterized by two parameters that can be chosen to be the
area $A$ (the ``size" of the black hole) and the angular momentum $J$. Moreover, these Kerr black
holes satisfy the well known lower bound $A\geq 8\pi|J|$. Fixing $|J|\neq 0$, then 
the smallest black hole satisfies $A=8\pi|J|$ and is called
extreme. In the realm of dynamical black holes, quantities like the quasi-local
angular momentum, do not have at the moment a clear parallel. However, axially symmetric black holes 
form a relevant class of dynamical black holes for which the quasi-local
angular momentum is, via Komar's formula,  well defined. Based on
heuristic physical arguments, it has been conjectured in \cite{dain10d} that
the same lower bound $A\geq 8\pi |J|$ indeed holds for dynamical axially symmetric black
holes. Interesting physical consequences on the evolution were discussed in \cite{dain10d}.  In \cite{Acena:2010ws}
\cite{Clement:2011kz} the bound was proved under several restrictive
assumptions and in the stationary case with matter and charge, it has been proved
in \cite{Hennig:2008zy}. Numerical evidence  was given in \cite{Jaramillo:2007mi}.  The purpose of this letter is to
present a general proof of this inequality for axially symmetric black
holes. Precisely, we extend the validity of the inequality from the unique
stationary Kerr black hole to all dynamical, in principle even very far from
equilibrium, dynamical axially symmetric vacuum black holes in the maximal gauge.  The proof provides also new
connections between black holes and stable minimal surfaces.

In order to describe the results we need to introduce some definitions.  An
initial data set of the Einstein vacuum equations, with cosmological constant
$\Lambda$, consists in a Riemannian three-manifold $S$ (possible with
boundary), together with its first and second fundamental forms, $h_{ij}$ and
$K_{ij}$ respectively, satisfying the constraints equations
\begin{equation}
\label{hamiltonianConst}
R+K^2-K_{ij}K^{ij} =2\Lambda,\quad
\nabla^iK_{ij}-\nabla_j K =0.
\end{equation}
In these equations, $K=h^{ij}K_{ij}$ and $R$ denotes the scalar curvature of
$h_{ij}$.  Initial data are called maximal if $K=0$.  The data are axially
symmetric if there exists a vector field with closed orbits $\eta^i$ such that
\begin{equation}
\label{eq:axial-sim}
\mathcal L_\eta h_{ij}=0,\qquad \mathcal L_\eta K_{ij}=0,
\end{equation}
where $\mathcal L$ denotes the Lie derivative.

For axially symmetric data the angular momentum $J$ associated to an arbitrary
 oriented closed surface $\Sigma$ in $S$ is defined by the surface integral
\begin{equation} 
  \label{eq:69} 
  J(\Sigma)=\int_\Sigma   \pi_{ij} \eta^i n^j \dvs, 
\end{equation} 
where $\pi_{ij}= K_{ij}- K h_{ij}$ and $n^i$, $\dvs$ are, respectively, the
unit normal vector and the area element of $\Sigma$. Note that $J$ represents
the angular momentum intrinsic to the surface $\Sigma$ and it depends only on
the homology class of $\Sigma$. It coincides with the total angular momentum of
an asymptotically flat end when the surface $\Sigma$ is homologous to an sphere
at infinity.

For axially symmetric data there exist two relevant scalars determined by the
Killing field: the square of its norm $\eta=\eta^i\eta^j h_{ij}$ and the twist
potential $\omega$, which can be computed in terms of second fundamental form
as follows (see \cite{Dain06c} for details).  Define the vectors $S^i$ and
$K^i$ by
\begin{equation}
 S_i=K_{ij}\eta^j-\eta^{-1}\eta_i K_{jk}\eta^j\eta^k,\quad
 K_i=\epsilon_{ijk}S^j\eta^k,  
\end{equation}
where $\epsilon_{ijk}$ is the volume element. Then, the momentum constraint
implies that the vector $K^i$ is locally a gradient
\begin{equation}
\label{eq:omega}
 K_i=\frac{1}{2}\nabla_i\omega.
\end{equation}
The twist potential $\omega$ evaluated at a surface $\Sigma$ determines its
angular momentum (see \cite{Acena:2010ws}).

We denote by $\gamma_{AB}$ and $\chi_{AB}$ the intrinsic metric and the second
fundamental form of a surface $\Sigma$.  The surface is called minimal if its
mean curvature (i.e. $\chi= \gamma^{AB}\chi_{AB}$) vanishes. A minimal surface
is called stable if it is a local minimum of the area. A surface is axially
symmetric if the Killing field $\eta^i$ is tangent to it.  For axially
symmetric surfaces, we have (outside the axis) a canonical adapted triad
defined by $(n^i, \xi^i, \eta^i)$, where $n^i$ is the unit normal vector to
$\Sigma$ and $\xi^i$ is an unit vector tangent to the surface and orthogonal to
$\eta^i$.

The local geometry near the horizon of an extreme Kerr black hole plays an
important role as limit case in our result. This geometry is characterized by
the concept of an extreme Kerr throat sphere, with angular momentum $J$,
defined as follows (see \cite{dain10d}). The sphere is embedded in an initial
data with intrinsic metric given by
\begin{equation}
\label{eq:gamma0}
\gamma_0=4J^2e^{-\sigma_0}d\theta^2+ e^{\sigma_0}\sin^2\theta d\phi^2,
\end{equation}
where 
\begin{equation}
\label{eq:sigma0}
\sigma_0=\ln(4|J|)-\ln(1+\cos^2\theta). 
\end{equation} 
Moreover, the sphere must be totally geodesic (i.e. $\chi_{AB}=0$), the twist
potential evaluated at the surface must be given by
\begin{equation}
\label{eq:omega0}
\omega_0=-\frac{8J\cos\theta}{1+\cos^2\theta},
\end{equation}
and  the  components of the second fundamental 
\begin{equation}
K_{ij}\xi^i= K_{ij}n^j n^i= K_{ij}\eta^j \eta^i = 0,
\end{equation}
must vanish at the surface.

The following is the main result of this article. 
\begin{theorem}
\label{t:main}
Consider an axisymmetric, vacuum and maximal initial data, with a non-negative
cosmological constant. Assume that the initial data contain
an orientable closed stable minimal axially symmetric surface
$\Sigma$. Then
\begin{equation}
\label{desigualdad}
 A \geq 8\pi |J|,
\end{equation}
where $A$ is the area and $J$ the angular momentum of $\Sigma$. Moreover, if
the equality in \eqref{desigualdad} holds then $\Lambda=0$ and the local
geometry of the surface $\Sigma$ is an extreme Kerr throat sphere.
\end{theorem}

Theorem \ref{t:main} has a remarkable consequence. Namely, for every orientable
and closed surface $\Sigma$ in a (complete) axisymmetric datum with several
asymptotically flat ends the inequality \eqref{desigualdad} holds.  It is in
particular satisfied by the event or marginal horizon of an axially symmetric
black hole. This proves the conjecture raised in \cite{dain10d}. We will
briefly outline this phenomenon. Further details will appear elsewhere.  For
such class of manifolds, it follows from a general result \cite{Meeks82}, that
for every closed surface $\Sigma$ there exist a finite set of
possibly repeated stable minimal surfaces $\{\Sigma_{i}\}$, such the sum of
its areas is equal to the infimum of the areas among all the isotopic variations
of $\Sigma$. Furthermore, because $\cup \Sigma_{i}$ is the measure theoretical limit of isotopies of 
$\Sigma$ (\cite{Meeks82}), it is deduced that  $|J(\Sigma)|\leq \sum |J(\Sigma^{o}_{i})|$ were
$\Sigma^{o}_{i}$ are those $\Sigma_{i}$'s that are orientable. Finally it is shown that, in our setting, each each $\Sigma^{o}_i$ must
be an axially symmetric sphere.  Theorem \ref{t:main} applies for each
$\Sigma_i$ and the claim follows.

\emph{Proof.}  --- We first observe that if $J\neq 0$ then the surface $\Sigma$
is diffeomorphic to $S^2$. This follows from a classical result of
\cite{Schoen79f} since the integral of the scalar curvature is strictly
positive on $\Sigma$. Let $F_t: \mathbb{R}\times S^2 \to S$ be a flow of
surfaces parametrized by $t\in\mathbb{R}$, such that $F|_{t=0}(S^2)=\Sigma$.
We impose that the family satisfies the equation $\dot F^i|_{t=0}=\alpha n^i$,
where dot denotes derivatives with respect to $t$, $n^i$ is the unit normal to
$\Sigma$ and $\alpha$ is an arbitrary function on $\Sigma$ that will be fixed
later on. As before, $\gamma_{AB}$ and $\chi_{AB}$ denote the intrinsic metrics
and the second fundamental forms of the surfaces $F_t(S^2)$.

The derivative of the mean curvature along the flow $F$ is given by 
\begin{equation}
\label{eq:evolchi}
\dot \chi = -\Lsig \alpha - (\chi_{AB}\chi^{AB}+R_{ij}n^in^j)\alpha,
\end{equation}
where $\Lsig$ is the Laplacian with respect to $\gamma_{AB}$. 

We use the relation
\begin{equation}
\Rs=R-2 R_{ij}n^in^j +\chi^2- \chi_{AB}\chi^{AB},
\end{equation}
to write $R_{ij}n^in^j$ in terms of $\Rs$ (the scalar curvature of
$\gamma_{AB}$) in equation \eqref{eq:evolchi}. We
obtain
\begin{equation}
\label{eq:evolchi2}
\dot \chi = -\Lsig \alpha - \frac{1}{2}(R-\Rs+\chi^2 +
\chi_{AB}\chi^{AB})\alpha. 
\end{equation}
For a minimal surface $\chi=0$ and the stability condition on $\Sigma$ implies
that
\begin{equation}
\label{eq:stability}
\ddot A|_{t=0}=\int \alpha \dot \chi \dvs\geq 0, 
\end{equation}
where $\dvs$ is the area element with respect to $\gamma_{AB}$. 

We multiply equation \eqref{eq:evolchi2} by $\alpha$, integrate it over
$\Sigma$ and use condition \eqref{eq:stability} to obtain
\begin{equation}
\label{eq:mainineq}
\int (|D\alpha|^2 +\frac{1}{2}\Rs \alpha^2)\dvs   \geq\frac{1}{2} 
\int \left(R + \chi_{AB}\chi^{AB}\right)
\alpha^2 \dvs. 
\end{equation} 
Note that  only derivatives intrinsic to $\Sigma$ appear on the
left hand side of this inequality.

By assumption, the surface $\Sigma$ is axially symmetric, therefore it
intersects the axis of symmetry at two points, which we define to be the poles
of $\Sigma$. A general axially symmetric metric on $S^2$ can be written in the
 form
\begin{equation}
\label{eq:coord}
\gamma=e^{\sigma}\left[e^{2q}d\theta^ 2+\sin^2\theta d\phi^2 \right],
\end{equation}
where $\sigma, q$ are regular functions of $\theta$. The coordinates
$(\theta,\phi)$ cover the sphere $\theta \in [0,\pi]$, $\phi \in [0,2\pi)$. The
poles are given by $\theta=0,\pi$. The axial Killing vector is given by
$\partial_\phi$ and the square of its norm is given by
\begin{equation}
\label{eq:etasigma}
\eta=e^{\sigma} \sin^2\theta.
\end{equation}
In these coordinates the determinant of the metric and the scalar curvature are
given respectively by
\begin{equation}
\sqrt{\det(\gamma)}=e^{\sigma+q}\sin\theta,
\end{equation}
\begin{equation}
\label{eq:Rs}
\Rs=\frac{e^{-\sigma-2q}}{\sin\theta}
\left(
2 q'\cos\theta+\sin\theta \sigma' q' +2\sin\theta -(\sin\theta \sigma')'
\right),
\end{equation}
where prime denotes derivative with respect to $\theta$. 

We want to find a change of coordinates $\tilde\theta(\theta)$ 
 such that in the new coordinates the metric has the same form, namely 
\begin{equation}
\label{eq:coordnew}
\gamma=e^{\tilde\sigma}\left[e^{2\tilde q}(d\tilde\theta^ 2)+\sin^2\tilde
  \theta d\phi^2\right]. 
\end{equation}
and such that
\begin{equation}
\label{eq:c}
\tilde \sigma + \tilde q=c,
\end{equation}
where $c$ is a constant. 
Comparing \eqref{eq:coord} with \eqref{eq:coordnew} we obtain the following
relations
\begin{equation}
e^{\sigma}\sin^2\theta=e^{\tilde\sigma}\sin^2\tilde\theta, \quad 
e^{\sigma/2+q}=e^{ \tilde\sigma/2+\tilde q} \tilde\theta'.
\end{equation}
 Using these equations and the condition \eqref{eq:c} we obtain
\begin{equation}
\tilde\theta' \sin\tilde\theta = e^{-c+\sigma+q}\sin \theta.
\end{equation}
This equation can be integrated to obtain 
\begin{equation}
\label{eq:tildetheta}
\cos\tilde\theta-1=-e^{-c}\int_0^\theta e^{\sigma+q}\sin \bar\theta
d\bar\theta. 
\end{equation}
Where we have fixed the integration constant with the condition $\tilde
\theta(0)=0$. The constant $c$ is fixed with the condition $\tilde
\theta(\pi)=\pi$. Using $\cos(\tilde\theta(\pi))=-1$, from
\eqref{eq:tildetheta} we obtain 
\begin{equation}
e^{c}=\frac{1}{2}\int_0^\pi e^{\sigma+q}\sin \bar\theta d\bar\theta.
\end{equation} 
The constant $c$ is related to  the area of the surface $\Sigma$ by
\begin{equation}
A=\int \dvs= 2\pi \int_0^\pi e^{\sigma+q}\sin \theta d\theta=4\pi e^{c}. 
\end{equation}
Note also that $\dvs=e^c\dv$, where $\dv=\sin\tilde\theta d\tilde\theta d\phi$
is the area element of the standard metric in $S^2$.
 
The regularity conditions on the metric at the axis imply that $\tilde
q(\tilde\theta=0,\pi)=0$. Hence, by equation \eqref{eq:c}, in these coordinates
we have
\begin{equation}
\label{eq:regularity}
\tilde\sigma(\tilde\theta=0)=\tilde\sigma(\tilde\theta=\pi).
\end{equation}
From now on, we assume that this coordinate system is used and we denote the
functions and the coordinates without the tilde.

The key step in the proof is to chose  the lapse function $\alpha$ to be
\begin{equation}
\alpha=e^{c-\sigma/2}.
\end{equation}
Using this choice of $\alpha$ we can explicitly calculate the left hand side of
inequality \eqref{eq:mainineq}
\begin{equation}
\int (|D\alpha|^2 +\frac{1}{2}\Rs \alpha^2 )\dvs=e^c \left(4\pi (c+1)-\int
(\sigma +\frac{1}{4} \sigma'^2) \dv \right),
\end{equation}
where we have used the expression \eqref{eq:Rs} for $\Rs$, the condition
\eqref{eq:c} and the boundary condition \eqref{eq:regularity}.  For the right
hand side of \eqref{eq:mainineq} we use the Hamiltonian constraint
\eqref{hamiltonianConst} and the hypothesis that the data are maximal to write
the scalar curvature as
\begin{equation}
R= K_{ij}K^{ij} +\Lambda.
\end{equation}
Using the adapted triad $(n^i,\xi^i,\eta^i)$, we write $K_{ij}K^{ij}$ as the
following sum of positive terms
\begin{multline}
\label{eq:KK}
K_{ij}K^{ij}= \left(K_{ij}n^in^j\right)^2+
\left(K_{ij}\xi^i\xi^j\right)^2 +
\eta^{-2}\left(K_{ij}\eta^i\eta^j\right)^2 +\\ 
2\left(K_{ij}\xi^in^j\right)^2 +2 \eta^{-1}\left(K_{ij}\eta^in^j\right)^2+
2\eta^{-1}\left(K_{ij}\eta^i\xi^j\right)^2.
\end{multline}
In \cite{Dain06c}, eq. (42), it has been proved that
\begin{equation}
  \left(K_{ij}\eta^i n^j\right)^2=
\frac{1}{4}\frac{\omega'^2}{\eta}e^{-\sigma-2q}.  
\end{equation}

Collecting these inequalities and discarding all the positive terms we obtain 
$8(c+1)\geq \mf$, where the important mass functional $\mathcal M$ (see
\cite{dain10d})  is defined by
\begin{equation}
\label{eq:4}
\mf =\frac{1}{2\pi}\int \left(\sigma'^2+
  4\sigma+\frac{\omega'^2}{\eta^2}\right) \dv.   
\end{equation}
Using the relation between $c$ and the area we finally obtain our main
inequality
\begin{equation}
A\geq 4\pi e^\frac{\mf-8}{8}.
\end{equation}
Inequality \eqref{desigualdad} follows from the bound 
\begin{equation}
  \label{eq:1}
 2|J|\leq e^{\frac{\mathcal M-8}{8}}.
\end{equation}
proved in  lemma 4.1 in \cite{Acena:2010ws} for all $\sigma,\omega$ such that
$\omega$ satisfies the boundary condition $\omega(0)=-\omega(\pi)=4J$
which ensures that $\Sigma$ has angular momentum $J$.  Note that Lemma 4.1 in \cite{Acena:2010ws} has a larger scope (not used or required here) 
as it applies to the extension of the functional (\ref{eq:4}) to non-axisymmetric functions $(\sigma,\omega)$.

It remains to prove the rigidity statement. We will prove that, if equality in
\eqref{eq:1} holds, then $\sigma=\sigma_0$ and $\omega=\omega_0$, where
$\sigma_0$ and $\omega_0$ are given by \eqref{eq:sigma0} and
\eqref{eq:omega0}. Having proved this, rigidity follows imposing $8\pi(c+1)={\mathcal{M}}$, and, using equation \eqref{eq:KK} in the now
equality \eqref{eq:mainineq}, track down all the null terms.

The strategy to prove rigidity is the following. If equality in \eqref{eq:1} is
achieved for the pair $(\sigma, \omega)$ then, being a minimum of $\mf$, it
must be a solution of the Euler-Lagrange equations of $\mf$. Interestingly, a
solution of the Euler-Lagrange equations of $\mf$ is also a solution of the
Euler-Lagrange equations of the functional 
\begin{equation}
\label{FSET}
\tilde{\mathcal{M}}_{\epsilon}=
\int_{\epsilon}^{\pi-\epsilon}\frac{\eta'^{2}+\omega'^{2}}{\eta^{2}}\sin\theta 
d\theta, 
\end{equation}
under smooth variations of compact support in $(\epsilon,\pi-\epsilon)$ for
every $\frac{\pi}{2}>\epsilon>0$ (see \cite{Acena:2010ws} for further
discussions). Where $\eta$ is given in terms of $\sigma$ by
\eqref{eq:etasigma}. Further, making the change of variables $\bar{s}=\ln
\tan\theta/2$ we have 
\begin{equation} 
\tilde{\mathcal{M}}_{\epsilon}=
\int_{\ln\tan
\epsilon/2}^{\ln\tan(\pi-\epsilon)/2}\frac{\eta'^{2}+
\omega'^{2}}{\eta^{2}}d\bar{s},  
\end{equation}
where prime in this equation denotes derivative with respect to $\bar s$.  It
is well known that a critical point of this functional (for every $\epsilon$
and under variations of compact support), namely solutions of its
Euler-Lagrange equations, are geodesics in the hyperbolic plane $
\mathbb{H}^2$. Here we are identifying the hyperbolic plane to the half plane
$\mathbb{R}^{2+}=\{(\eta,\omega)/\eta>0\}$ together with the (hyperbolic
metric) $(d\eta^{2}+d\omega^{2})\eta^{-2}$. The geodesics are parametrized by
$\bar{s}$ where $\bar{s}$ and the arc-length are related by
$s=c_{1}\bar{s}+c_{2}$ where $c_{1},c_{2}$ are constants. Thus, the pair
$(\eta(\theta), \omega(\theta))$ representing our minimizing solution will be a
geodesic $\gamma(\bar{s}(\theta))$ in the hyperbolic plane. Now, geodesics of $
\mathbb{H}^2$, are either half circles or half lines perpendicular to the axis
$\{\eta=0\}$. The boundary condition $\omega(0)=-\omega(\pi)=4J$ on $\omega$ fixes
the geodesic to be a centered half circle with radius fixed by the angular
momentum and equal to $4|J|$. The only freedom left is thus that of the
parametrization. This freedom, as shown below, is fixed using
\eqref{eq:regularity}. The solution found in this way will be unique and
extreme Kerr.

Following the line of reasoning described, we will compute explicitly the
solution using a complex expression for geodesics in the hyperbolic plane. In
complex notation $\gamma=\omega+i\eta$, it is $ \gamma=\frac{ae^{s}i+b}{ce^{s}
i+d}$. One has (suppose $a/c>0$ but the analysis is the same otherwise)
$4|J|=\omega(\pi)=a/c$ and $-4|J|=\omega(0)=b/d$ (note $c\neq 0$ and $d\neq
0$). Thus
\begin{equation}
\label{SOL} 
\eta= \textnormal{Im}(\gamma)=
\frac{cd(\frac{a}{c}-\frac{b}{d})e^{s}}{c^{2}e^{2s}+d^{2}}
=\frac{8(\frac{c}{d})|J|e^{s}}{(\frac{c}{d})^{2}e^{2s}+1}.
\end{equation} 
To find the general solution in terms of $\theta$ we need to find $c_{1}$
using the Euler-Lagrange equations of (\ref{FSET})
\begin{equation}
\left(\frac{\omega'\sin\theta}{\eta^{2}}\right)'=0,\quad \left(\sin\theta
\frac{\eta'}{\eta}\right)'+\frac{\omega'^{2}}{\eta^{2}}\sin\theta=0.
\end{equation}
The first equation implies
$\omega'=\lambda_{0}\frac{\eta^{2}}{\sin^{2}\theta}$, where $\lambda_{0}$ is a
constant and therefore
$(\frac{\omega'}{\eta})^{2}=\lambda_{0}^{2}(\frac{\eta}{\sin\theta})^{2}$.
Inserting $\omega'$ into the second equation, multiplying it by
$(\eta'\sin\theta)\eta^{-1}$ and, finally, integrating it in $\theta$, brings
us to the identity $\sin^{2}\theta
(\eta'/\eta)^{2}+\lambda_{0}^{2}\eta^{2}=\lambda_{1}$, where $\lambda_{1}$ is a
constant. From this identity, the expression
$\eta'/\eta=\sigma'+2\frac{\cos\theta}{\sin\theta}$ and the fact that $\sigma$
is bounded at the poles we deduce that $\lambda_{1}=4$. Thus
$(\eta'^{2}+\omega'^{2})\eta^{-2}=4\sin^{-2}\theta$. It
follows that $c_{1}=2$ and $s=c_{2}+\ln\tan^{2}\frac{\theta}{2}$. The general
solution of $\eta$ is found making $\frac{c}{d}e^{c_{2}}=\beta$ in (\ref{SOL}),
explicitly
\begin{equation} 
\eta=\ln 4|J|+\ln
\frac{2\beta\tan^{2}\frac{\theta}{2}}{\beta^{2}\tan^{2}\frac{\theta}{2}+1},
\quad \beta>0.  
\end{equation}
The condition $\sigma(0)=\sigma(\pi)$ implies $\beta=1$. In this case a
 trigonometric manipulation shows that $\eta=\ln 4|J|-\ln 1+\cos^{2}\theta$
 which is the expression for $\eta$ of extreme Kerr.

\emph{Final remarks.} --- As shown in \cite{Chrusciel:2011eu}, any
axially symmetric sphere has an adapted coordinate system as required in
\cite{Acena:2010ws} to deduce inequality \eqref{desigualdad}.  It may seem
thus, that part of Theorem \ref{t:main} could be derived from
\cite{Acena:2010ws} without further elaborations. In fact, the condition $\dot \chi \geq0$ 
in \cite{Acena:2010ws} can be
replaced by $\int \alpha \dot \chi \dv\geq 0 $, which is similar to \eqref{eq:stability} but has different area element. 
It is not clear a priori how these two integral
inequalities can be reciprocally implied. Our approach avoids the use
of special coordinates.

\begin{acknowledgments}
S. D. would like to thank Mu-Tao Wang for discussions. 
This work took place during the visit of S. D. to the Max Planck Institute for
Gravitational Physics in 2011.  He thanks for the hospitality  of
this institution.  S. D. is supported by CONICET (Argentina). This work was
supported in part by grant PIP 6354/05 of CONICET (Argentina), grant Secyt-UNC
(Argentina) and the Partner Group grant of the Max Planck Institute for
Gravitational Physics (Germany).
\end{acknowledgments}

%\bibliography{/home/dain/biblio/biblio}

\begin{thebibliography}{9}
\expandafter\ifx\csname natexlab\endcsname\relax\def\natexlab#1{#1}\fi
\expandafter\ifx\csname bibnamefont\endcsname\relax
  \def\bibnamefont#1{#1}\fi
\expandafter\ifx\csname bibfnamefont\endcsname\relax
  \def\bibfnamefont#1{#1}\fi
\expandafter\ifx\csname citenamefont\endcsname\relax
  \def\citenamefont#1{#1}\fi
\expandafter\ifx\csname url\endcsname\relax
  \def\url#1{\texttt{#1}}\fi
\expandafter\ifx\csname urlprefix\endcsname\relax\def\urlprefix{URL }\fi
\providecommand{\bibinfo}[2]{#2}
\providecommand{\eprint}[2][]{\url{#2}}

\bibitem[{\citenamefont{Dain}(2010)}]{dain10d}
\bibinfo{author}{\bibfnamefont{S.}~\bibnamefont{Dain}}, \bibinfo{journal}{Phys.
  Rev. D} \textbf{\bibinfo{volume}{82}}, \bibinfo{pages}{104010}
  (\bibinfo{year}{2010}).

\bibitem[{\citenamefont{Aceña et~al.}(2011)\citenamefont{Aceña, Dain, and
  Cl\'ement}}]{Acena:2010ws}
\bibinfo{author}{\bibfnamefont{A.}~\bibnamefont{Ace\~na}},
  \bibinfo{author}{\bibfnamefont{S.}~\bibnamefont{Dain}}, \bibnamefont{and}
  \bibinfo{author}{\bibfnamefont{M.~E.~G.} \bibnamefont{Cl\'ement}},
  \bibinfo{journal}{Classical and Quantum Gravity}
  \textbf{\bibinfo{volume}{28}}, \bibinfo{pages}{105014}
  (\bibinfo{year}{2011}).

\bibitem[{\citenamefont{Gabach~Cl\'ement}(2011)}]{Clement:2011kz}
\bibinfo{author}{\bibfnamefont{M.~E.} \bibnamefont{Gabach~Cl\'ement}}
  (\bibinfo{year}{2011}), \eprint{1102.3834}.

\bibitem[{\citenamefont{Hennig et~al.}(2010)\citenamefont{Hennig, Cederbaum,
  and Ansorg}}]{Hennig:2008zy}
\bibinfo{author}{\bibfnamefont{J.}~\bibnamefont{Hennig}},
  \bibinfo{author}{\bibfnamefont{C.}~\bibnamefont{Cederbaum}},
  \bibnamefont{and} \bibinfo{author}{\bibfnamefont{M.}~\bibnamefont{Ansorg}},
  \bibinfo{journal}{Commun. Math. Phys.} \textbf{\bibinfo{volume}{293}},
  \bibinfo{pages}{449} (\bibinfo{year}{2010}).

\bibitem[{\citenamefont{Jaramillo et~al.}(2007)\citenamefont{Jaramillo, Vasset,
  and Ansorg}}]{Jaramillo:2007mi}
\bibinfo{author}{\bibfnamefont{J.~L.} \bibnamefont{Jaramillo}},
  \bibinfo{author}{\bibfnamefont{N.}~\bibnamefont{Vasset}}, \bibnamefont{and}
  \bibinfo{author}{\bibfnamefont{M.}~\bibnamefont{Ansorg}}
  (\bibinfo{year}{2007}),  \eprint{0712.1741}.

\bibitem[{\citenamefont{Dain}(2008)}]{Dain06c}
\bibinfo{author}{\bibfnamefont{S.}~\bibnamefont{Dain}}, \bibinfo{journal}{J.
  Differential Geometry} \textbf{\bibinfo{volume}{79}}, \bibinfo{pages}{33}
  (\bibinfo{year}{2008}).

\bibitem[{\citenamefont{Meeks et~al.}(1982)\citenamefont{Meeks, Simon, and
  Yau}}]{Meeks82}
\bibinfo{author}{\bibfnamefont{W.}~\bibnamefont{Meeks}, \bibfnamefont{III}},
  \bibinfo{author}{\bibfnamefont{L.}~\bibnamefont{Simon}}, \bibnamefont{and}
  \bibinfo{author}{\bibfnamefont{S.~T.} \bibnamefont{Yau}},
  \bibinfo{journal}{Ann. of Math. (2)} \textbf{\bibinfo{volume}{116}},
  \bibinfo{pages}{621} (\bibinfo{year}{1982}).

\bibitem[{\citenamefont{Schoen and Yau}(1979)}]{Schoen79f}
\bibinfo{author}{\bibfnamefont{R.}~\bibnamefont{Schoen}} \bibnamefont{and}
  \bibinfo{author}{\bibfnamefont{S.-T.} \bibnamefont{Yau}},
  \bibinfo{journal}{The Annals of Mathematics} \textbf{\bibinfo{volume}{110}},
  \bibinfo{pages}{127} (\bibinfo{year}{1979}).

\bibitem[{\citenamefont{Chrusciel and Nguyen}(2011)}]{Chrusciel:2011eu}
\bibinfo{author}{\bibfnamefont{P.~T.} \bibnamefont{Chrusciel}}
  \bibnamefont{and} \bibinfo{author}{\bibfnamefont{L.}~\bibnamefont{Nguyen}},
  \bibinfo{journal}{Class.Quant.Grav.} \textbf{\bibinfo{volume}{28}},
  \bibinfo{pages}{125001} (\bibinfo{year}{2011}).

\end{thebibliography}

\end{document}